 \documentclass[final,5p,times,twocolumn]{elsarticle}

\usepackage{amssymb}

\journal{Astroparticle Physics}

\begin{document}

\begin{frontmatter}



\title{Cosmic ray acceleration in magnetic circumstellar bubbles}


\author{V.N.Zirakashvili, V.S.Ptuskin}

\address{Pushkov Institute of Terrestrial Magnetism, Ionosphere and Radiowave
Propagation, 108840, Troitsk, Moscow, Russia}

\begin{abstract}
We consider the diffusive shock acceleration in interstellar bubbles
created by powerful stellar winds of supernova progenitors. Under
the moderate stellar wind magnetization the bubbles are filled by
the  strongly magnetized low density gas. It is shown that the
maximum energy of
 particles accelerated in this environment can exceed the "knee" energy in the observable cosmic ray spectrum.

\end{abstract}


\end{frontmatter}



\section{Introduction}

 It is well established that supernova remnants (SNRs)
are efficient accelerators of protons, nuclei and electrons. They are the principle sources
of Galactic cosmic rays. The diffusive shock acceleration  (DSA) mechanism
 \cite{krymsky77,bell78,axford77,blandford78} can provide the acceleration
of the charged particles at shocks of SNRs. During the last decades
the excellent results of X-ray and gamma-ray astronomy supplied
the observational evidence of the presence of multi-TeV energetic
particles in these objects (see e.g. \cite{lemoine14} for a review).

It is clear now that some magnetic amplification mechanism is needed for the acceleration of
 cosmic ray protons beyond PeV energies. The most popular one is the amplification in the course
 of the non-resonant streaming instability suggested by Bell \cite{bell04}. The instability is produced by
 the electric current of highest energy particles escaping to the upstream region of the quasiparallel shock. In this regard the accelerated
particles prepare the magnetic inhomogeneities for
effective DSA themselves. It was shown recently that the similar instability
 operates at quasiperpendicular shocks \cite{matthews17}.

The amplified magnetic fields indeed present in young SNRs as it was
determined from the thickness of X-ray
 filaments \cite{voelk05}. The highest magnetic fields were found for young SNRs Cas A and Tycho. One can expect that
 the particle acceleration is very efficient in these objects.
However the recent gamma ray spectral measurements performed for
these remnants revealed spectral breaks or even
 cut-offs at TeV energy \cite{kumar15,park15,guberman17,archambaul17}.  This means that at present the protons are
accelerated to energies of the order of 10 TeV in these remnants.
The maximum energies were not significantly higher in the past because
of rather high shock speeds about 5 thousands km s$^{-1}$ at present. The
most probable explanation is the small size of magnetic disturbances
generated via the non resonant streaming instability and the
suppression
 of the effective acceleration by large scale electric fields appearing under the development of this instability \cite{zirakashvili15}.

On the other hand there are several more efficient cosmic ray accelerators like RXJ1713.7-3946, Vela Junior with maximum energies  close to 100 TeV
 and without the strong magnetic amplification \cite{lemoine14}.
Probably the blast wave  in these remnants propagates in the low
density ($n<0.1$ cm$^{-3}$) medium produced by the stellar wind of
the supernova progenitor. So the unexpected results of the modern
gamma-ray astronomy draw our attention to DSA in such environments.

In this paper we consider DSA in wind blown cavities. As we shall show below the magnetic field in cavities produced by powerful winds of
 O-stars and Wolf-Rayet (WR) stars can be high enough to provide the acceleration of cosmic ray particles up to PeV energies.

The paper is organized as follows. In the next Sections 2-4 we
describe the magnetic structure and the particle acceleration in wind blown bubbles.
The numeric results of magnetohydrodynamic (MHD) simulations and DSA modeling are  given in Sections 5 and 6 respectively.
The discussion of results and conclusions are given in Sections 7 and 8.

\section{Acceleration in wind blown bubbles}

V\"olk and Biermann suggested acceleration of cosmic rays in stellar wind cavities \cite{voelk88}. As the magnetic field is
almost azimuthal at large distances from the progenitor star the spherical shock wave produced in the supernova explosion is quasiperpendicular.
Acceleration rate at such shocks can be higher than the rate obtained in the so called Bohm limit when the scattering
frequency of particles $\nu $ is comparable with the gyrofrequency $\Omega $ \cite{jokipii87,takamoto15,giacalone17}.

However the energy gain at the perpendicular shock is accompanied by
the drift of particles along the shock surface in the direction
perpendicular to the regular magnetic field. The particles get
energy in the electric field ${\bf E}=-[{\bf u}\times {\bf B}]/c$ in
the shock frame. The maximum energy is determined by the electric potential difference under the condition $\nu \ll \Omega $.
For acceleration in the stellar wind
the resulting maximum energy is comparable with the one obtained
 in the Bohm limit. It is important that the high level of MHD turbulence in the shock vicinity is not necessary for the efficient acceleration
  in this quasiperpendicular regime.  The low level of background turbulence that presents in the stellar wind can be enough.
This is contrary to the quasiparallel shocks where the high level of MHD turbulence is needed for the justification of the Bohm diffusion.

We can estimate
 the maximum energy  $E_{\max }$ of particles using the relation $D_B(E_{\max })=\kappa V_sR_s$, where $R_s$ and $V_s$ are the
 shock radius and speed respectively, $D_B(E)=\mathrm{v}^2/3\left| \Omega \right| $ is the Bohm diffusion coefficient 
of particles with velocity {v},
$\kappa \sim 0.03-0.3$ is the numeric factor (see the Sect. 6 below). This factor $\kappa =1/3$
 if the maximum energy is
 limited by the electric potential mentioned above.

For the steady
 stellar wind with the mass loss rate $\dot{M}$ and the wind speed $u_w$ the gas density $\rho $ and the magnetic field strength $B$  at large distances $r$
 from the star are given by

\begin{equation}
\rho =\frac {\dot{M}}{4\pi u_wr^2}, \ B=\pm \sin \theta \frac {B_s\Omega _sr_s^2}{u_wr}=\pm \sin \theta u_w\frac {\sqrt{4\pi \rho }}{M_w}.
\end{equation}
Here $B_s$, $\Omega _s$ and $r_s$ are the radial component of
the magnetic field, the rotation rate and the radius of the star respectively
and $\theta $ is the colatitude. It is more convenient to use the
magnetosonic Mach number of the wind  $\left. M_w=u_w\sqrt{4\pi \rho }/B\right| _{\theta =\pi /2}$ below. It is related
with a frequently used magnetization parameter $\sigma $ that is the
ratio of magnetic and
 kinetic energy densities of the flow as $\sigma =M_w^{-2}$.
The maximum energy of particles with charge $q$ is then

\[
E_{\max }^w=3\kappa qBV_sR_s/c=\frac {3\kappa }{M_w}\frac{qV_s}{c}\sqrt{u_w\dot{M}}=
\]
\begin{equation}
70\ Z\ \mathrm{PeV}\frac {3\kappa }{M_w}\frac{V_s}{c}\left( \frac {\dot{M}}{10^{-5}M_{\odot } \mathrm{yr}^{-1}} \right) ^{1/2}
\left( \frac {u_w}{10^{3}\ \mbox{km\  s}^{-1} }\right) ^{1/2}
\end{equation}
Here $Z$ is the charge number and we use the wind parameters of WR star. Then for $M_w=20$, $\kappa =0.1$
and the shock speed $V_s=10^4$ km s$^{-1}$ that is
a characteristic value for Ib/c supernovae with the ejecta mass $M_{\mathrm{ej}}=M_{\odot }$ and the energy
of explosion $E_{\mathrm{SN}}=10^{51}$ erg  \cite{chevalier05}
we  obtain the maximum proton energy 35 TeV.

The stellar wind is bounded by the termination shock at distance $r=R_{TS}$ where the magnetic field strength and the gas density increase
by a factor of $\sigma _{TS}$, where $\sigma _{TS}\approx 4$
is the shock compression ratio. The gas flow is almost incompressible  downstream of the shock and the gas velocity $u$ drops as $r^{-2}$.
The azimuthal magnetic field increases linearly with the distance $r$ in this region \cite{cranfill71, axford72, chevalier92,zhang96}.
This is a so called Cranfill effect \cite{nerney91}.
At distances where the magnetic energy is comparable
with the gas pressure magnetic stresses begin to influence the gas flow.  We can use
 the energy conservation along the lines of the flow for the description of this effect

\begin{equation}
\frac {u^2}{2}+\frac {\gamma }{\gamma -1}\frac {P}{\rho }+\frac {B^2}{4\pi \rho }=\mathrm{const}.
\end{equation}

In addition the gas density and azimuthal magnetic field strength are related as $\rho r\sin \theta /B =\mathrm{const}$.  The first term in Eq. (3) can be neglected in the incompressible flow. Then the gas pressure and magnetic field strength can be
 found as functions of distance $r$ in the equatorial plane $\theta =\pi /2$ (see papers \cite{axford72,kennel84,begelman92,chevalier94} for details).

At large distances the magnetic energy dominates thermal and kinetic energies
 and drops as $r^{-2}$ similar to the supersonic wind region but with an additional
amplification factor $M_w^2/2$ of the magnetic field \cite{chevalier92}. The same is true for the maximum energy of particles accelerated when the blast wave
 propagates in the downstream region of the termination shock. However the system should have the size large enough for this. We can
write down these two regimes in one expression for the maximum
energy of particles accelerated by the blast wave in the downstream
region that is in the wind blown bubble

\begin{equation}
E_{\max }^b=E_{\max }^w\min \left( \frac {M_w^2}2, \sigma _{TS}\frac {R_s^2}{R_{TS}^2}
\right)
\end{equation}
where the maximum energy in the wind zone $E_{\max }^w$ is given by
Eq. (2). So even the modest ratio $R_s/R_{TS} \sim 3$ results in
 acceleration to PeV energies for blast waves of Ib/c supernovae.

Note that we have a well known physical object to check our estimate
given by Eq. (2). Anomalous cosmic rays that are single charged ions
are accelerated up to hundreds MeV at the termination shock of the
solar wind \cite{cummings87}. Using the solar wind parameters
$V_s=u_w=400$ km s$^{-1}$, $\dot{M}=2.5\cdot 10^{-14}$ $M_{\odot }$
yr$^{-1}$, $M_w=20$ and $\kappa =1/3$ we get $E_{\max }^w= 150$ MeV.

We can use the theory of wind blown bubbles \cite{weaver77} to estimate the radius of the termination shock in Eq. (4). Rewriting
Eq. (12) from Weaver et al. \cite{weaver77}

\begin{equation}
\frac {R_c}{R_{TS}}=\left( \frac {25}{44}\right) ^{1/2}
\left( \frac {15}{16}\right) ^{3/4}\sqrt{\frac {u_wt}{R_c}}
\end{equation}
and taking into account that the radius of the contact discontinuity
$R_c$ is close to the bubble radius $R_b$ we obtain

\[
\frac {R_b}{R_{TS}}=\left( \frac {25}{44}\right) ^{1/2}
\left( \frac {15}{16}\right) ^{3/4}\left( \frac {308\pi }{125}\right) ^{1/10}
u_w^{0.3} t^{0.2}\dot{M}^{-0.1}\rho _0^{0.1}
\]
\begin{equation}
\approx t_\mathrm{kyr}^{0.2}n_0^{0.1}\left( \frac {\dot{M}}{10^{-5}M_{\odot } \mathrm{yr}^{-1}} \right) ^{-0.1}
\left( \frac {u_w}{10^{3}\ \mbox{km\  s}^{-1} }\right) ^{0.3}
\end{equation}
where $\rho _0$ and $n_0$ are the mass and number densities of the surrounding medium and we use Eq. (21) of Weaver et al. \cite{weaver77}
for the bubble radius $R_b$:

\[
R_b=\left( \frac {125}{308\pi }\right) ^{0.2}
u_w^{0.4} t^{0.6}\dot{M}^{0.2}\rho _0^{-0.2}=
\]
\begin{equation}
 0.52\ \mathrm{pc}\ t_\mathrm{kyr}^{0.6}n_0^{-0.2}\left( \frac {\dot{M}}{10^{-5}M_{\odot } \mathrm{yr}^{-1}} \right) ^{0.2}
\left( \frac {u_w}{10^{3}\ \mbox{km\  s}^{-1} }\right) ^{0.4}
\end{equation}

For the bubble age $t_{\mathrm{kyr}}\sim 300$ that is the expected duration of WR stage and for the density $n_0=10$ cm$^{-3}$
in the parent molecular cloud
the termination shock
radius is a factor of 4 smaller than the
bubble radius $R_b=9$ pc. Then  the last factor in Eq. (4) is of the order of 60 when the blast wave approaches the bubble boundary
and the maximum energy of protons is several PeV for SNRs of Ib/c supernova. Note that the blast wave has swept up 3$M_{\odot }$ of
gas at this time and therefore
 is in the transition to the Sedov stage when the shock contains most of the explosion energy.

We assumed that the the stellar wind is the only source of the gas in the bubble. Contrary to this assumption
Weaver at al. \cite{weaver77} took into account the
 evaporation of the gas from the bubble shell. The evaporation was regulated by the thermal conductivity flux from the
hot interior to the cold shell. In the real situation the conductivity can be suppressed by the azimuthal magnetic field
 and
 by the scattering of thermal electrons on magnetic perturbations in the turbulent medium of the bubble.

In the opposite case considered by Weaver et al. \cite{weaver77} the gas density in the bubble is significantly higher
in comparison with
our estimates.
Then the blast wave quickly decelerated after the crossing of the termination shock and particles can not be accelerated
to PeV energies.
This result was already obtained by Berezhko and V\"olk in their modeling of DSA in wind blown bubbles \cite{berezhko00}.

For convenience we give expressions for the gas density and magnetic
field strength
 in the bubble below. Using Eq, (16) of Weaver et al. \cite{weaver77} we get

\[
\rho =\frac {25}{11}\frac {\dot{M}t}{4\pi R_b^3}
\left( 1-\frac {r^3}{R_b^3}\right) ^{-8/33}=
 8.6\cdot 10^{-25}\frac {\mathrm{g}}{\mathrm{cm}^3}\\ t_\mathrm{kyr}^{-0.8}n_0^{0.6}
\]
\begin{equation}
\left( \frac {\dot{M}}{10^{-5}M_{\odot } \mathrm{yr}^{-1}} \right) ^{0.4}
\left( \frac {u_w}{10^{3}\ \mbox{km\  s}^{-1} }\right) ^{-1.2}
\left( 1-\frac {r^3}{R_b^3}\right) ^{-8/33}
\end{equation}

For the magnetic field strength at the equator $\theta =\pi /2$ we obtain

\[
B =\frac {25}{11}\frac {\sqrt{\dot{M}u_w^3}t}{M_wR_b^2}
\frac {r}{R_b}\left( 1-\frac {r^3}{R_b^3}\right) ^{-8/33}=
\frac {690\mu \mathrm{G}}{M_w}\ t_\mathrm{kyr}^{-0.2}n_0^{0.4}
\]
\begin{equation}
\left( \frac {\dot{M}}{10^{-5}M_{\odot } \mathrm{yr}^{-1}} \right) ^{0.1}
\left( \frac {u_w}{10^{3}\ \mbox{km\  s}^{-1} }\right) ^{0.7}
\frac {r}{R_b}\left( 1-\frac {r^3}{R_b^3}\right) ^{-8/33}
\end{equation}

Eqs (8) and (9) give infinite values of density and magnetic field at the bubble boundary. This means that the approximation
 used is not good enough in this region. However this happens only in the narrow region near the bubble boundary. That is why we shall
omit the last factor in Eqs (8) and (9) for estimates below.

For $M_w=20$, $t_{\mathrm{kyr}}\sim 300$ and for the density $n_0=10$ cm$^{-3}$ we obtain the number density $n_H=0.015$ cm$^{-3}$ and the magnetic
 field strength $B=30\ \mu $G at the periphery of  the bubble.

We used only WR stage of stellar evolution to estimate the  bubble parameters. We would obtain larger bubble size taking
into account the main sequence phase. However this situation  is less probable because of the possible motion of the star
relative to the circumstellar gas. For example the motion with $10$ km s$^{-1}$ will leave far behind the star almost all low density gas
produced during several
 millions years of the main sequence stage. This is not so for the shorter and more energetic WR phase (see also the modeling below).

\section{Injection problem}

It is known that the injection of ions is suppressed at quasiperpendicular shocks \cite{caprioli14}. However there are several possibilities
to avoid this problem for the case of supernova explosion in the stellar wind.

The magnetic field of stellar winds is radial close to the star. Therefore the supernova shock is quasiparallel in the very beginning. Then the injection of
 ions is efficient during the short time after the  explosion. The accelerated particles will generate waves and turbulence via streaming instability.
The amplified magnetic field is almost isotropic and therefore
injection will take place at quasiparallel parts of the shock even
when the blast wave will reach large distances where undisturbed
magnetic field is azimuthal. Probably this situation is  stable when
the shock is modified by the pressure of accelerated particles. The
modified shocks are self regulated systems when decrease (increase)
of injection results in stronger (weaker) subshock and in the
corresponding spectral flattening (steepening) at suprathermal
energies that compensates initial decrease (increase) of injection.
Some observational support for this scenario is given by
 observations of radio-supernovae of Type Ib/c. The radio-spectra of these supernovae are steep with radio-index -1 \cite{chevalier06},
  that corresponds to $E^{-3}$
 spectrum of electrons. Such sub-GeV energetic electrons are accelerated at the subshock with compression ratio 2.5, that is the shock is modified. The density of
 WR winds where the blast wave of Ib/c supernovae propagates is relatively low and synchrotron losses of radioelectrons in the amplified
magnetic field are not strong enough to produce the spectral steepening \cite{chevalier06}.

Another place where the injection can take place is the downstream
region of the stellar wind termination shock. It is expected that
for high Mach numbers of the wind weak density disturbances in the
wind will produce vortex motions and the corresponding magnetic
amplification in the downstream region \cite{giacalone07}. When the
blast wave crosses the termination shock, the injection will take
place at quasiparallel parts of the shock because of the chaotic
 orientations of amplified random magnetic fields.
The azimuthal and latitudinal components of the random magnetic field will be further amplified in the incompressible
 radial flow similar to the regular field. This effect is important for weak regular fields that is for
 high Mach numbers $M_w$.

It is also possible that quasiparallel parts of the shock can appear
after the interaction with dense clumps in the bubble. Such clumps can
be produced when the fast WR wind disrupts the shell of the Red Super
Giant (RSG) gas  ejected at previous stages of the progenitor
evolution (see hydrodynamical modeling \cite{veelen09}).

We conclude that probably the ion injection takes place in the turbulent medium of the bubble.


On the other hand it is known that the electron injection indeed takes place at quasiperpendicular shocks \cite{requelme11} even
without  the ion injection. In this regard our model of DSA in the bubbles gives some theoretical support to the leptonic models
of gamma emission in SNRs.

\section{Influence of magnetic fields on the rotation of hot stars}

Up to now kG magnetic fields were detected for several O-stars by direct
measurements of Zeeman effect \cite{donati09}. WR stellar winds  also reveal their magnetic fields via
 observed nonthermal X-ray and radio emission \cite{becker07}. The Mach number of the wind $M_w$ can
be written in terms of the surface magnetic field $B_s$ as (see Eq. (1))
\[
M_w=\frac {u_w}{\Omega _sr_s} \frac {\sqrt{\dot{M}u_w}}{B_sr_s}=
2.5\frac {u_w}{\Omega _sr_s}
\left( \frac {B_s}{100\ \mathrm{G}}\right) ^{-1}
\]
\begin{equation}
\times
\left( \frac {\dot{M}}{10^{-5}M_{\odot } \mathrm{yr}^{-1}} \right) ^{1/2}
\left( \frac {u_w}{10^{3}\ \mbox{km\  s}^{-1} }\right) ^{1/2}
\left( \frac {r_s}{10^{12}\ \mathrm{cm}}\right) ^{-1}.
\end{equation}

The first factor in this expression is the ratio of the wind speed $u_w$ and the rotation speed $v_\mathrm{rot}=\Omega _sr_s$ of the star.
For rapidly rotating young massive stars it is of the order of 10. It is clear from Eq. (10) that the fast rotation of the supernova
 progenitor is a necessary condition for acceleration to PeV energies.

Therefore we should check whether the strong magnetic torque
influences the star rotation. The loss of the angular momentum $J$
according to the theory of axisymmetric MHD flows (e.g.
\cite{zirakashvili96}) is given by

\begin{equation}
\frac {dJ}{dt}=-\frac 23\dot{M}\Omega _sr_a^2
\end{equation}
Where $r_a$ is the distance to the Alfv\'en point in the spherically symmetric wind outflow. The factor $2/3$ in this equation
 comes from the integration on colatitude.

For the sake of simplicity we shall assume that the stellar density is uniform at $r<r_s$. Then the stellar
angular momentum $J=0.4M_sr_s^2\Omega_s$ where $M_s$ is the mass of the star. Eq. (11) can be rewritten as

\begin{equation}
 M_s\frac {d\Omega_s}{dt}=\frac 53\Omega _s\left( \frac {r_a^2}{r_s^2}-1\right) \frac {dM_s}{dt}
\end{equation}

For the wind velocity profile

\begin{equation}
u(r)=u_w\sqrt{1-\frac {r_s}{r}}
\end{equation}
that is expected for the wind driven by radiation the position of the Alfv\'en point can be found from the equation
\begin{equation}
\frac {r_a^2}{r_s^2}\sqrt{1-\frac {r_s}{r_a}}=\frac {B_s^2r_s^2}{\dot{M}u_w}=\frac {v_a^2}{v^2_\mathrm{rot}}
\end{equation}
where $v_a=u_w/M_w$ is the Alfv\'en velocity at large distances in the stellar wind. So the evolution
 of the stellar rotation is governed by the ratio ${v_a}/{v_\mathrm{rot}}$. The solution of Eq. (12) for the time independent
 ${v_a}/{v_\mathrm{rot}}$ is

\begin{equation}
\Omega _s=\Omega _{s0}\left( \frac {M_s}{M_{s0}}\right) ^{\frac53\left( \frac {r_a^2}{r_s^2}-1\right) }.
\end{equation}
Where $\Omega _{s0}$ and $M _{s0}$ are the initial angular velocity and the mass of the star respectively.

It is clear from this solution that for the case $r_a\gg r_s$ the weak change of the angular velocity is possible only for the
 weak change of the stellar mass during the stellar evolution (e.g. for the Sun).  For the massive stars the change of the mass during
 the stellar evolution is significant. Then the weak change of the angular velocity is possible only for $r_a\approx r_s$. We can rewrite then
Eq. (15) using the  approximate solution of Eq. (14) as

\begin{equation}
\Omega _s=\Omega _{s0}\left( \frac {M_s}{M_{s0}}\right) ^{\frac{10}3\frac{v_a^4}{v_\mathrm{rot}^4}}.
\end{equation}

It is clear that the ratio ${v_a}/{v_\mathrm{rot}}\leq 0.5$ is enough for weak changes of the angular velocity. The corresponding Mach number of the
 wind

\begin{equation}
M_w\geq 2\frac {u_w}{v_\mathrm{rot}}.
\end{equation}
For the fast stellar rotation $u_w/v_\mathrm{rot}=10$ it is enough to have the Mach number $M_w\geq 20$ for the weak change of
the angular velocity during the stellar evolution. Interestingly this value $M_w\sim 20-30$ is enough for the acceleration to PeV energies (see Sect.2).

\begin{figure}
\includegraphics[width=9.0cm]{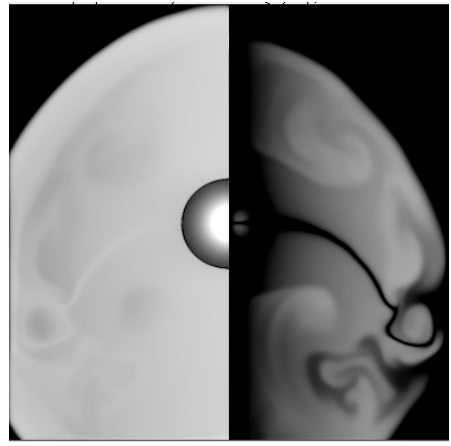}
\caption{ Distribution of the gas pressure $P_g$ (left panel) and the magnetic tension $B^2/4\pi $  (right panel)
in the domain 10$\times $20 pc at $t=3\cdot 10^5$ yr. The logarithmic scaling is from $2.3\cdot 10^{-12}$ erg cm$^{-3}$ (black color)
to $2.3\cdot 10^{-10}$ erg cm$^{-3}$ (white color). }
\end{figure}

\begin{figure}
\includegraphics[width=8.0cm]{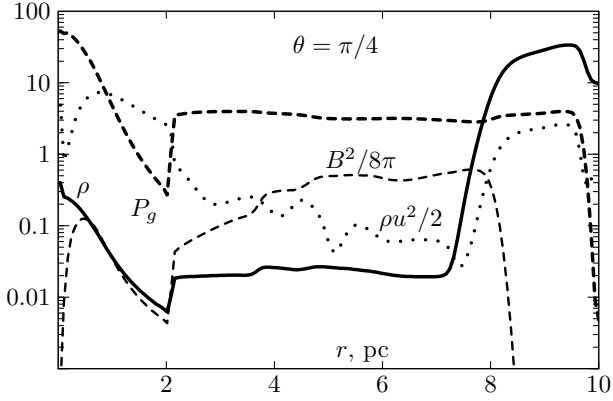}
\caption{ Dependence of the gas density $\rho $, gas pressure $P_g$,
magnetic pressure $B^2/8\pi $, and kinetic energy density $\rho
u^2/2$ on distance $r$  at $\theta =\pi /4$. The gas, magnetic
pressure and energy density are in units  $2.3\cdot 10^{-11}$ erg
cm$^{-3}$, the gas density  is in units $2.3\cdot 10^{-24}$ g
cm$^{-3}$. }
\end{figure}

\section{MHD modeling of WR bubble}

We performed two dimensional MHD modeling of the bubble created by WR star. We used the Total Variation Diminishing hydrodynamic scheme
\cite{trac03} with the Van Leer flux limiter. The method was applied in cylindrical
 coordinates and the azimuthal magnetic field was added as an additional hydrodynamic quantity. The radiative cooling was not taken into
 account. It is important for the formation of the high density thin outer shell of the bubble. However the acceleration of cosmic rays at the forward shock
of the bubble  and the interstellar magnetic fields can strongly
influence the shell structure. We were interested in the inner
part of the bubble and the pure MHD modeling is good enough for this
purpose.

The numeric results obtained for
 the grid with 200$\times $400 cells and size 10$\times $20 pc are shown below.
The number density of circumstellar medium is $n_0=10$ cm$^{-3}$.
The supersonic stellar wind outflow was obtained adding the sources of the gas, magnetic field and energy in the central region.
We used the wind speed 1000 km s$^{-1}$, mass loss rate $\dot{M}=10^{-5}M_{\odot } \mathrm{yr}^{-1}$ and the Mach number $M_w=20$.
To model the motion of the
 central star we fix the negative value $u_z=-10$ km s$^{-1}$ at the upper boundary of the simulation domain. The results at the end of simulation at
$t=3\cdot 10^5$ yr are shown in Fig. 1 and 2.

\begin{figure}
\includegraphics[width=8.0cm]{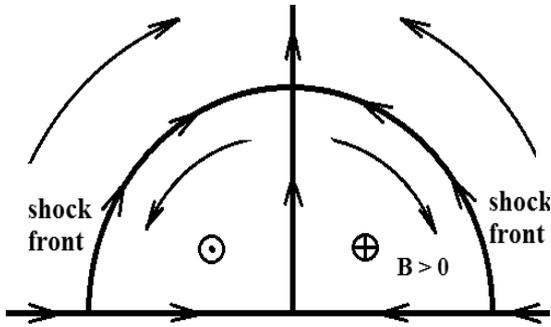}
\caption{ Drift motions of the protons accelerated at the shock propagating in the stellar wind. }
\end{figure}

It is clear that the bubble periphery is filled by a rather strong magnetic field with strength about 20 $\mu $G. The magnetic  pressure is
comparable with the gas pressure. The magnetic fields of opposite polarity are separated by the thin neutral current sheet that is influenced by
 vortex motions of the gas.
\begin{figure}
\includegraphics[width=8.0cm]{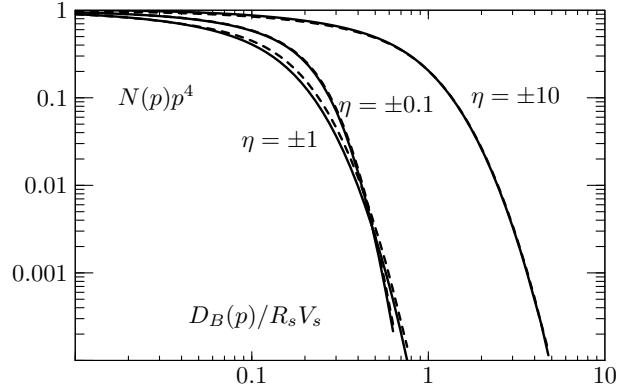}
\caption{ Averaged on the colatitude spectra  of particles at the shock propagating in the stellar wind. The spectra for positive and
 negative values of $\eta $ are shown by the solid and dashed curves respectively.  }
\end{figure}

\begin{figure}
\includegraphics[width=8.0cm]{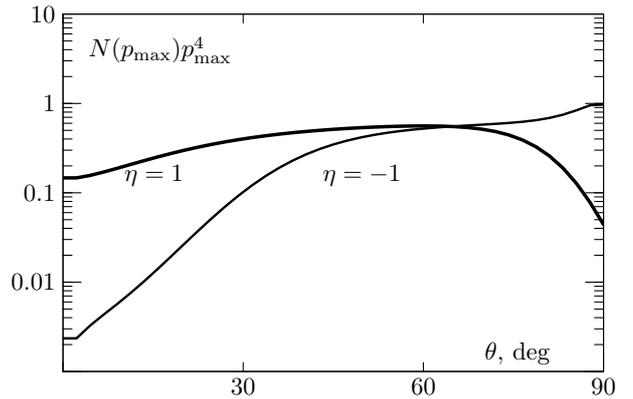}
\caption{ Distribution of accelerated particles with momentum $p=p_{\max }$  on colatitude $\theta $ at the shock
for the case $\eta =1$ (thick curve) and
 $\eta =-1$ (thin curve). }
\end{figure}

\section{Modeling of DSA at perpendicular shocks}

The particle acceleration at the solar wind termination shock in the presence of diffusion and drifts was considered 
 by Jokipii \cite{jokipii86}. 
We performed the similar modeling of DSA at the spherical nonmodified  perpendicular shock with the radius $R_s(t)=V_st$ and
the compression ratio 4 moving with the constant speed $V_s$.

The cosmic ray transport equation for the isotropic momentum distribution $N({\bf r},t,p)$ has the form \cite{berezinsky90}

\begin{equation}
\frac{\partial N}{\partial t} + \textbf {u}\nabla N - \nabla _iD_{ij}\nabla _jN 
-\frac {\nabla \textbf {u}}{3} p\frac{\partial N}{\partial p} = Q
\end{equation}
Here ${\bf u}$ is the gas velocity and $Q$ is the source term. 


The diffusion tensor  $D_{ij}$ in Eq. (18) can be presented as the sum of the symmetric and antisymmetric parts 
\begin{equation}
D_{ij}=(D_{\parallel }-D_{\perp })b_ib_j+D_{\perp }\delta
_{ij}+D_Ae_{ijk}b_k, 
\end{equation}
where ${\bf b}={\bf B}_0/|{\bf B}_0|$  is the  unit vector in the direction of the regular magnetic field 
${\bf B}_0$, and   $D_{\parallel }$, $D_{\perp }$,  $D_A$  are the parallel, perpendicular and antisymmetric diffusion 
 coefficients respectively. Introducing the drift velocity ${\bf u}_d=[\nabla \times D_A{\bf b}$] we can rewrite the transport equation as 

\[
\frac{\partial N}{\partial t} + (\textbf {u+u}_d)\nabla N  
-\frac {\nabla \textbf {u}}{3} p\frac{\partial N}{\partial p} = 
\]
\begin{equation}
 \nabla ({\bf b}(D_{\parallel }-D_{\perp })({\bf b}\nabla N)) +\nabla D_{\perp }\nabla N+Q
\end{equation}

We use the hard sphere scattering \cite{dolginov67} below. The diffusion coefficients have the following form 
\begin{equation}
D_{\parallel }=\frac {\mathrm{v}^2}{3\nu }, \ D_{\perp }=\frac {\mathrm{v}^2\nu
/3}{\Omega ^2+\nu ^2}, \ D_{A}=\frac {\mathrm{v}^2|\Omega |/3}{\Omega ^2+\nu
^2}.
\end{equation} 

We shall assume that the parameter $\eta =\nu /\Omega $  that is  the ratio  of the scattering frequency
$\nu $ and gyrofrequency $\Omega $   does not depend on coordinates. The cosmic ray transport equation 
can be rewritten in new variables $\xi =r/R_s(t)$,
 colatitude $\theta $ and $\tau =\ln (R_s(t)/R_0)$ in the upstream region of the perpendicular shock $\xi >1$ as

\[
\frac {\partial N}{\partial \tau}-\xi \frac {\partial N}{\partial \xi }= \frac {d(p,\tau )}{\sin \theta  }
\left( \frac {1}{\xi ^2}\frac {\partial }{\partial \xi }\xi ^{2-n_1}\frac {\partial N}{\partial \xi }+\right.
\]
\begin{equation}
\left. \xi ^{-n_1-2}\frac {\partial ^2N}{\partial \theta ^2}
+\xi ^{-n_1-2}\frac {1-n_1}{\eta }\frac {\partial N}{\partial \theta }
\right)
\end{equation}
where  the dimensionless function
$d(p,\tau )=D_{\perp }(p)/R_sV_s$ is determined by the perpendicular
diffusion coefficient $D_{\perp }=D_B|\eta |/(1+\eta ^2)$ at the
shock position. It was assumed that the magnetic field strength is a
power law function on $\xi $ that is $B\sim \xi ^{n_1}\sin \theta $.
The last term in Eq. (22) describes the latitudinal drift of
 particles in the nonuniform magnetic field. As  for the drift velocity in the  radial direction it is not zero only at the poles and at the
equator where the magnetic field changes the sign (see Fig.3). This was taken into account by boundary conditions (see Eq. (25) below).

In the downstream region $\xi <1$ we used the following equation.
\[
\frac {\partial N}{\partial \tau}-\frac {\xi }4 \frac {\partial N}{\partial \xi } -\frac {3p}4 \frac {\partial N}{\partial p }=\frac {d(p,\tau )}{4\sin \theta  }
\left( \frac {1}{\xi ^2}\frac {\partial }{\partial \xi }\xi ^{2-n_2}\frac {\partial N}{\partial \xi }+\right.
\]
\begin{equation}
\left.
\xi ^{-n_2-2}\frac {\partial ^2N}{\partial \theta ^2}
+\xi ^{-n_2-2}\frac {1-n_2}{\eta }\frac {\partial N}{\partial \theta }
\right)
\end{equation}
Here we assumed the linear profile of the radial gas velocity $u(\xi )=3V_s\xi /4$ at $\xi <1$.
This results in additional adiabatic energy losses of particles in
the downstream region. It was also assumed that the magnetic field
strength increases by a factor of 4 in the shock transition region.

\begin{figure}
\includegraphics[width=8.0cm]{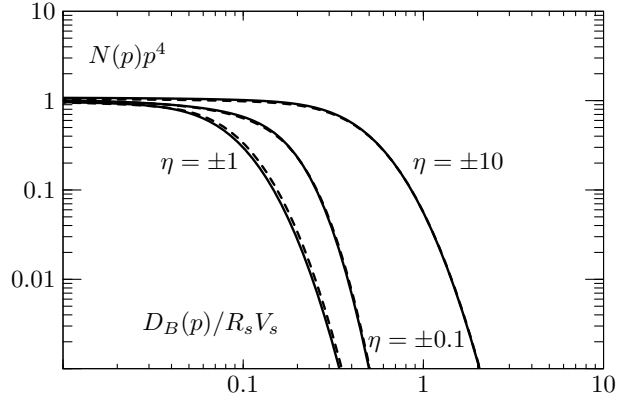}
\caption{ Averaged on the colatitude spectra  of particles at the shock propagating in the bubble. The spectra for positive and
 negative values of $\eta $ are shown by the solid and dashed curves respectively.}
\end{figure}

\begin{table}
\begin{center}
\caption{Values of $\kappa $ for different values of $\eta $}
\begin{tabular}{|c|c|c|c|c|c|c|}
\hline $\eta$  & $\pm $0.01 & $\pm $0.1 & -1.0 & 1.0 & -10 &  10\\
\hline wind    &  0.18& 0.14& 0.10 &0.09 & 0.59& 0.58 \\
\hline bubble  & 0.18 & 0.15& 0.09 &0.08 & 0.48& 0.48 \\
\hline
\end{tabular}
\label{tab:k}
\end{center}
\end{table}

The boundary condition at the shock front at $\xi =1$ is given by
\[
\frac {d(p,\tau )}{\sin \theta } \left( \left. 4\frac {\partial
N}{\partial \xi }\right|_{\xi =1+0}-\left. \frac {\partial
N}{\partial \xi }\right|_{\xi =1-0} \right) =
\]
\begin{equation}
= p\frac {\partial N_R}{\partial p}-\frac {3d(p)}{\eta \sin \theta
}\frac {\partial N_R}{\partial \theta }-4Q_R(p)
\end{equation}
Here $N_R=N|_{\xi =1}$ is the momentum distribution at the shock
front and the source term $Q_R(p)$ describes the injection of low
energy particles at the shock front. The  boundary condition at $\theta =0$ and $\theta
=\pi /2$ is

\begin{equation}
\xi \frac {\partial N}{\partial \xi }= \eta \frac {\partial N}{\partial \theta }
\end{equation}

We shall consider two important cases below. If the shock propagates in the wind  the function $d(p)$ is time independent and the index $n_1=-1$.
The linear profile of the gas speed in the
 downstream region results in $n_2=2$. In the second case the shock propagates in the bubble with the linear increase of the magnetic field strength.
The maximum energy of particles increases and the function $d(p,\tau )$ contains an additional factor $\exp (-2\tau )$.
The corresponding numbers are $n_1=1$ and $n_2=10$. The magnetic field is concentrated in the narrow region behind the shock in this case.

Small energy particles were injected uniformly at the shock surface. The transport equations (22), (23) with boundary
conditions (24), (25) were solved using the finite difference method.
The temporary evolution was calculated up to $\tau =1$ when the
shock radius
 had increased $e$ times in comparison with the initial radius $R_0$. For the wind case the momentum distribution is almost steady at this time.

The numeric results are shown in Fig. 4-6 and in Table 1.


Spectra of particles at the shock propagating in the wind are shown
in Fig. 4. In spite of opposite directions of drifts the spectra
integrated on colatitude are almost the same for different signs of
$\eta $. For large absolute values of $\eta $ this is expected
because in this case the diffusion is almost isotropic and  the
regular field plays no r\^ole. At small absolute values of $\eta $
the particles are magnetized and their energy gain is accompanied by
 the drift on colatitude  at the shock surface (see the right hand side of the boundary condition (24)). The maximum energy is regulated by the electric potential
 difference only. The spectra are slightly different in the intermediary case $\eta \sim 1$. On the other hand the latitudinal distribution of highest energy
particles
 strongly depends on the sigh  of $\eta $ (see Fig.5). For $\eta <0$ the particles drift to equatorial regions of the shock and their number density
 is maximal at $\theta =\pi /2$. For the opposite sign the particles drift to the pole but this effect is compensated by weak magnetic fields in the polar
region.

The values of the factor $\kappa $ in Eq. (2) were obtained in our modeling and are given in Table 1.
The maximum momentum $p_{\max }$ was determined as
 $N(p_{\max })p_{\max }^4=e^{-1}$. For small absolute values of $|\eta |<0.1$ the factor $\kappa $ tends to the universal value $\kappa =0.19$.

The values  $\eta \sim 0.1-1$ are important for DSA in the bubbles. We expect that these values of $\eta $  are realized in the turbulent
 medium with magnetic perturbations which are comparable with the regular magnetic field. This situation is also favorable for the ion injection
at the blast wave
 propagating in the bubble. The results of our modeling correspond to the age of the blast wave $t=R_s/V_s$. This age is $t\sim 10^3$ yr 
 when the blast wave of Ib/c supernova explosion approaches the bubble boundary. 


\section{Discussion}

We found that wind blown bubbles produced by fast stellar  winds of the supernova progenitors are very plausible places for DSA to PeV energies.
The magnetic field of the
 stellar wind is significantly amplified in the bubble by the Cranfill effect (see Sect. 2). The magnetic pressure is comparable with the gas pressure at the periphery of
 the bubble. A significant part of the stellar wind energy goes in to the magnetic energy.
In addition to the regular magnetic field there are magnetic
 disturbances in the turbulent medium of the bubble.  In this regard the stellar wind prepares ideal
conditions for DSA at the blast wave of the supernova explosion.

Therefore we expect that there are two types of young SNRs as sources of Galactic cosmic rays.

The first one includes
the young SNRs where the particles
 accelerated at the the blast wave prepare the conditions for their efficient acceleration via the streaming instability resulting in
the generation of MHD turbulence and in the magnetic
 field amplification in the
 upstream region of the shock. SNRs produced by supernova explosions in the interstellar medium and in the circumstellar medium with
 a low level of background turbulence or weak regular magnetic fields are of this type. The examples are SN 1006, Tycho, Cas A.
The maximum energy is of the order of 100 TeV for this type of SNRs.

The second type includes the young SNRs in the wind blown bubbles where the low density medium is turbulent and is prepared for DSA via the magnetic
amplification by the Cranfill effect.  The examples are RX J1713.7-3946, RCW 86, Vela Jr.
The maximum energy can be close to PeV or even more for remnants of Ib/c supernovae.

This picture is in accordance with radio, X-ray and gamma ray
observations of young SNRs.

The similar conclusion about the efficient cosmic ray acceleration by supernova blast wave propagating in stellar winds of WR and O-stars
 and the prediction of  two kinds of cosmic ray accelerating SNRs was made earlier by Biermann \cite{biermann93}. Later his model was
used in several phenomenological models of Galactic cosmic rays
\cite{zatsepin06,thoudam16}.

Our consideration of DSA in wind blown bubbles is close to the model of Berezhko and V\"olk \cite{berezhko00}. A new element of our model is
 the magnetic amplification produced by the Cranfill effect.
In addition we neglect
the gas evaporation from the bubble shell (see Sect. 2). This results in the lower bubble density, the higher shock speed and in the proton
acceleration beyond PeV energies. Our opinion is that the use of the gas evaporation based on the classical thermal conductivity
is questionable in the turbulent magnetized plasma.

The maximum energies close to 0.5 PeV were obtained by Telezhinsky et al. \cite{telezhinsky13}
in their modeling of DSA in remnants of Ic supernovae. They did not take into account the Cranfill effect  and used the stronger
magnetic field  of the stellar wind. In addition a rather high mass of supernova ejecta $M_{\mathrm{ej}}=5M_{\odot}$ was assumed. The lower ejecta mass
 $M_{\mathrm{ej}}\sim 1M_{\odot }$ would  result in acceleration to PeV energies.

The regular magnetic field of several tens of $\mu $G in the bubble is comparable with
the random field generated in the upstream region of the shock by
 streaming instability of accelerated particles. At first sight this means that the maximum energy is not higher than the maximum energy  provided by
 the streaming instability. However this is not so because the random field is concentrated in the narrow region near the shock. In addition the magnetic
field of the bubble is azimuthal and this results in the more effective
confinement of accelerated particles.

It is expected that the strong magnetic fields of the bubble prevent the strong shock modification. Then the spectra of
 accelerated particles are not very hard. Contrary to the case of quasiparallel shocks this will not result in lower maximum energies because the maximum
 energies provided by quasiperpendicular shocks are not related with the number density of highest energy accelerated particles.

Another important point is that the Pevatrons in the bubbles are long lived accelerators
with the age of thousands years. The WR stars are the progenitors
 of Type Ib/c supernova which are $\sim 10\% $ of all core collapse supernovae.
Then for the core collapse supernova rate $0.02$ yr$^{-1}$  we expect to find
 several Pevatrons in the remnants of Type Ib/c supernova with the thousand year age.

This situation is opposite to
 the probable acceleration to PeV energies in high density environment of remnants produced in the Type IIn and IIb supernova explosions.

Because of the low densities
the expected fluxes of hadronic gamma rays
 are of the order of $10^{-12} n_0^{0.6}d_{\mathrm{kpc}}^{-2}$ erg cm$^{-2}$ s$^{-1}$ when the blast wave propagates inside the bubble.
The high number density  $n_0>10$ cm${^{-3}}$ of the gas around the bubble is preferable for the gamma detection.
Probably several such objects were already detected by
 the modern Cherenkov telescopes (e.g. the resent HESS detection of RX J1741-302 \cite{Hess17}).

Parameters of the WR bubble obtained in our MHD modeling
correspond to the environment of SNR RX J1713.7-3946. In the
leptonic model of gamma emission weak magnetic fields $\sim 15\ \mu
$G in the remnant shell \cite{aharonian06} correspond to the Mach number $M_w\sim 100$
of the stellar wind.

The Mach number is not constrained in the
hadronic model because the interaction with the dense shell of the
bubble is necessary to reproduce the observable flux of gamma
emission.
 Deceleration of the blast wave in the shell can result in the modest current maximum energy
of the order of 100 TeV
derived for hadronic scenario \cite{aharonian06}. It is possible that protons were accelerated to higher energies at earlier times when the blast wave
propagated inside the bubble.

Strong regular magnetic fields of the bubble can manifest themselves
  via Faraday rotation with rotation measure RM$\sim 10$ rad m$^{-2}$ and via high level of the linear polarization of SNR radio emission.
In this regard Pevatrons in magnetic bubbles combine
 the features of young and old SNRs. It is expected that they emit nonthermal X-rays like young SNRs and  highly polarized radio emission
 with magnetic fields vectors tangent to the remnant rims like in old SNRs.

\section{Conclusions}

Our main conclusions are the following:

1) Powerful stellar winds of O-stars and WR stars can produce strongly magnetized interstellar bubbles with
the magnetic energy comparable with the thermal
 energy of the hot low density gas inside the bubbles.

2) We expect that there are two types of SNRs as the sources of Galactic cosmic rays. The first one includes the SNRs produced by
supernova explosions in the interstellar medium or in the circumstellar medium with weak turbulence and weak regular magnetic fields.
The second
 type includes SNRs in wind blown bubbles where the low density medium is prepared for the efficient DSA. Both types of SNRs are observed now in radio,
X-rays and gamma rays.

3) The estimated maximum energy of protons accelerated in
remnants of Ib/c supernova produced by WR winds is well beyond PeV energy. It is slightly below PeV but not too much for remnants of
Type IIP supernovae produced by
 winds of O-stars.

4) Since the lack of hydrogen is a characteristic feature of WR stellar winds, we explain the helium dominated cosmic ray
composition in the "knee" region.

5) We expect to find several Galactic Pevatrons with age $\sim 10^3$ years in magnetic bubbles contrary to the previous estimates.

We thank Heinz V\"olk for interesting discussions of DSA in SNRs.
The work was partially supported by the Russian Foundation for Basic Research grant 16-02-00255.












\end{document}